# Elastic wave imaging with Maxwell's fish-eye lens

Liuxian Zhao[1,*], Chunlin Li[2], Xuxu Zhuang[1], Hao Guo[1], Yongquan Liu[2, *]

[1]Institute of Sound and Vibration Research, Hefei University of Technology, 193 Tunxi Road, Hefei 230009, China

[2]State Key Laboratory for Strength and Vibration of Mechanical Structures, Department of Engineering Mechanics, School of Aerospace Engineering, Xi'an Jiaotong University, Xi'an 710049, China

*Author to whom correspondence should be addressed: lxzhao@hfut.edu.cn and liuy2018@xjtu.edu.cn

**ABSTRACT**

In this paper, a modified Maxwell's fish-eye lens is proposed in order to achieve super-resolution imaging. This lens possesses elevated refractive index profile compared with the traditional Maxwell's fish-eye lens. The refractive index profile is achieved with variable thickness configuration defined in a sheet plate structure, to realise desired changes in refractive indices. The wave propagation behaviours and the full width at half maximum (FWHM) are obtained from numerical simulations and experimental studies at the focal region of the lens. Super-resolution imaging is observed in a broadband frequency scope, with the FWHM around 0.2λ from 5 to 10 kHz. This work provides a straightforward and flexible approach to the engineering of the MFEL imaging characteristics and energy distributions for related applications.

# 1. Introduction

Achieving effective wave focusing and improving image quality has become a prominent topic in recent decades due to its relevance for academic research and engineering applications. One example is the use of ultrasound focus technology in medical applications. Extracorporeal ultrasound is focused on the location of the tumour within the body, and then the accompanying energy in the ultrasound waves is concentrated, ultimately inhibiting tumour growth [1-3]. In addition, focusing acoustic waves is also significant for harvesting energy and reducing vibration, because it facilitates energy harvesting after the wave is focused at a fixed point [4-10].

The phenomenon of negative refraction, based on the negative group velocity in phononic crystals (PCs) and metamaterials, has been developed to meet these requirements [11-14]. However, due to transmittance limitations, most of the energy has been reflected because impedance mismatch, and the transmitted energy at the focus position is very weak. A gradient index lens, with a refractive index profile that changes according to a hyperbolic secant function, is capable of redirecting the normal incident plane waves [15-18]. Most gradient-index lens are formulated utilizing phononic crystals, with the focal point situated within the medium. Simultaneously, it can be fashioned to concentrate the incoming wave by employing a meta-surface that synthetically regulates the wave phase [19]. On the other hand, several other lenses including the Luneburg lens [20-23] and the Maxwell's fish-eye lens (MFEL) [24, 25] are designed and manufactured. For example, Zhao *et al*. [26] proposed a novel gradient index structural lens based on the concept of generalized Luneburg lens for the realization of double foci. The results demonstrated that ultralong subwavelength focusing can be achieved for a broadband frequency range. It should be noted that all of the aforementioned techniques concentrate mechanical vibration energy on a focal area that has a full width at half maximum (FWHM) of approximately half the wavelength. This is due to limitations caused by Rayleigh diffraction. In general, due to the basic diffraction limitation imposed upon elastic waves, the resulting focusing size is constrained to the range of (0.5λ, λ), where λ refers to the elastic wavelength [27].

Smaller focus sizes, indicating higher energy concentration, are anticipated for energy harvesting or collection, non-destructive testing, and medical imaging. An increasing number of lenses based on PCs and metamaterials are being developed and examined to enhance the quality of focusing or imaging and surpass the diffraction limit [27]. Accordingly, some mechanisms, such as the unattenuated transport of evanescent modes from the source across the lens and the conversion of evanescent components into propagation waveforms [28], can be identified and used. For example, studies have shown that hyperbolic metamaterials with extremely anisotropic densities or stiffnesses can effectively transmit subwavelength information from the source to the output limit, enabling subwavelength imaging [29]. Fuentes-Domínguez *et al*. [30] employed additive manufacturing technique to print a circular array of micropillars on an aluminium slab for surface acoustic waves focusing based on the theory of acoustic Luneburg lens. The lens is obtained by exploiting the dispersion properties of the metamaterials arising from the well-known resonant coupling between the micropillars, which shows a strong subwavelength wave focusing. Nonetheless, for most of the methods mentioned above, a fixed frequency or a specific frequency range with a relatively small bandwidth is applicable. In contrast to the previous works, Chen *et al*. [31] designed a flexural wave Mikaelian lens for high resolution focusing through igniting evanescent waves at the interface of conformally mapped Mikaelian lens with hyperbolic secant refractive index profile. After that, Li *et al*. [32] proposed a new method to improve the focusing of reflected waves based on the division of a metasurface lens into two adjacent lenses having inner angles. Compared to a single lens, the focal position will capture a larger number of reflected waves from broader angles, resulting in an enhanced focus size. Unfortunately, due to design scheme, the proposed lens exhibits a poor omni-direction.

Inspired by the pioneering work of acoustic super-resolution imaging of airborne waves, which is built upon a solid immersion 3D Maxwell's fish-eye lens [33], the solid immersion technique is introduced to the elastic waves on a thin plate structure for super-resolution imaging in this work. Compared with the previous lens, the Maxwell's fish-eye lens works in a broadband frequency range and omni-directionally. Furthermore, this study achieves a sub-wavelength focusing phenomenon, whereby the focal size is approximately $0.2\lambda$. At first, the scheme of design is clarified, and the COMSOL Multiphysics software is used to examine working performance. Then, experimental measurements for the modified Maxwell's fish-eye lens (MMFEL) are performed to further prove the correctness of the methodology. The suggested design methodology is anticipated to guide the exploration of physical mechanisms and the development of structures for ultimate wave focusing usefulness. The novelty of this study is that the refractive index distribution of a traditional MFEL is modified based on the solid immersion method. The MMFEL is designed using variable thickness of thin plate structure. The super-resolution imaging of elastic waves based on the MMFEL can be realized in a wide frequency range for the first time.

## 2. Design of The Modified Maxwell Fish-Eye lens

Figure 1 presents the schematic diagram of optical imaging of two ideal points A and B through a conventional optical lens. In the object space, the spacing between the two points A and B is $d_0$; the angle between the chief rays of the two points is $\theta$; the object distance is $l_0$; the refractive index is $n$. In the image space, the spacing between the two image A' and B' is $d_i$; the angle between the chief rays of the two image points is $\theta'$; the image distance is $l_i$; the size of the point-spread-function is $d$; the refractive index is $n'$. The radius of the lens is $R$. Generally, when the refractive indices are equal, *i.e.* $n=n'$, the angles in both sides of the lens are also equal, *i.e.* $\theta=\theta'$.

According to the Rayleigh diffraction limit, the achievable minimum size of the point-spread function on the image plane is given by $d = 0.61\lambda/NA$, where $NA=n_i \cdot \sin(\operatorname{atan}(R/l_i))$ with $n_i$ is the refractive index of the lens, and the two points A and B cannot be distinguished when the spacing between the two image spots is less than $d$, as shown in Figure 1. Therefore, the corresponding angular resolution in the object space is restricted to $\delta\theta_{DL} = 2\operatorname{atan}(0.5d/l_i)$, which can be rewritten as equation (1). If $l_i$ is much greater than $R$, it can be further simplified as equation (2).

$$\delta\theta_{DL}(NA) = 2\operatorname{atan}(0.5 \times 0.61\lambda/(NA \cdot l_i)) \qquad (1)$$

$$\delta\theta_{DL}(0) = 1.22\lambda/D \qquad (2)$$

where $D$ is the diameter of the lens ($D=2R$). It can be seen from equations (1) and (2) that the two images can be resolved or distinguished when the separation of the two acoustic sources is greater than $\delta\theta_{DL}$ in Figure 1, while the image pattern is similar to that for a single acoustic source when the separation of the two acoustic sources is smaller than $\delta\theta_{DL}$.

In a more general case, if the two angles are not equal and $\theta'$ is greater than $\theta$, or $\theta'=K\theta$ ($K > 1$), then the angular resolution $\delta\theta_K$ in the object space should be modified by adding an additional factor $1/K$ to equation (1), as expressed in equation (3). Similarly, it can be simplified as equation (4) when $l_i$ is much greater than $R$.

$$\delta\theta_K(NA) = 2\operatorname{atan}(0.5 \times 0.61\lambda/(K \cdot NA \cdot l_i)) \qquad (3)$$

$$\delta\theta_K(0) = 1.22\lambda/(K \cdot D) \qquad (4)$$

Therefore, the corresponding resolution can be enhanced by a factor of $K$. Such angular magnification has been realized for airborne acoustic waves [33].

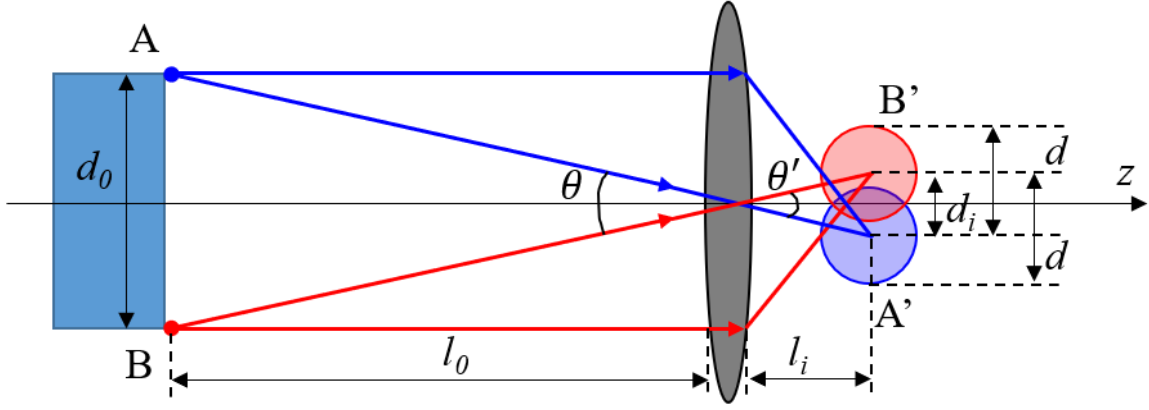

**Figure 1: Schematic of acoustic imaging.**

Consider a spherical symmetry system as shown in Figure 2(a), which consists of a spherical object with a refractive index profile as follows [34]:

$$n(r) = \begin{cases} \dfrac{2n_0}{1+\left(\dfrac{r}{R_0}\right)^2} & (0 \leq r < R_0) \\ 1 & (r > R_0) \end{cases} \quad (5)$$

in which $n_0$ is the elevated refractive index compared with the background medium whose refractive index is 1, $r$ is the distance from the centre of the lens ($r = \sqrt{x^2 + y^2}$), and $R_0$ is the radius of the lens.

Note that the refractive index between the lens and the background medium is not consistent, inducing acoustic impedance mismatching at the interface and causing reflections from the interface. Therefore, a transition zone is introduced at the periphery of the lens, so as to prevent significant interface reflection, hence disturbing the imaging as little as possible. The refractive index distribution of the MMFEL is given below:

$$n(r) = \begin{cases} \dfrac{2n_0}{1+\left(\dfrac{r}{R_0}\right)^2} & (0 \leq r < R_1) \\ 1 & (r > R_1) \end{cases} \quad (6)$$

where $R_1 = \sqrt{2n_0 - 1}R_0$, and the area between $R_0$ and $R_1$ is the transition zone, as shown in Figure 2(b).

For flexural waves propagating through a plate structure with variable thickness, the phase velocity $c_p$ can be calculated as a function of thickness $h$ using the expression $c_p = (\frac{\omega^2 h^2 E}{12(1-\nu^2)\rho})^{\frac{1}{4}}$. The parameters $\omega$, $\rho$, $E$ and $\nu$ represent the angular frequency, mass density, Young's modulus, and Poisson ratio of the thin plate, respectively. By Snell's law, $n = \frac{c_0}{c_p}$, in which $c_0$ is the phase velocity of flexural wave propagating through structure of constant thickness $h_0$. The refractive index distribution as a function of plate thickness is then exported $n = \sqrt{\frac{h_0}{h}}$ [20]. By combining this equation for the refractive index with Equation (6), we obtain Equation (7) which is plotted in Figure 2(c).

$$h(r) = \begin{cases} \frac{[1+(r/R_0)^2]^2}{4n_0^2} h_0 & (0 \leq r < R_1) \\ 1 & (r > R_1) \end{cases} \quad (7)$$

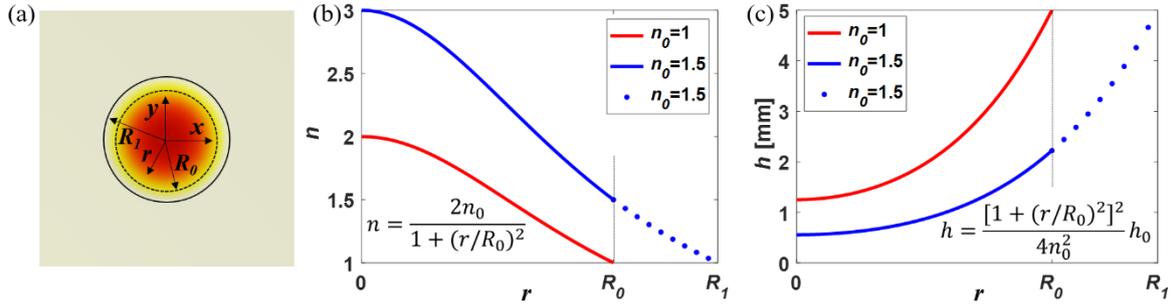

**Figure 2: (a) The distribution of refractive index of MMFEL. (b) The relationship of refractive index with the radial distance. (c) The relation of variable thickness with the radial distance.**

The thickness profile of the MMFEL is shown in Figure 3(a) can be obtained, from Equation (7). In this study, we choose $R_0$ = 100 mm, $h_0$ = 5 mm and $n_0$ = 1.5 in the following analyses. The dimension of the thin plate structure is 650 ×650 mm, as provided in Figure 3 (a). For comparison, a traditional Maxwell's fish-eye lens (MFEL) is investigated with $n_0$ = 1, as provided in Figure 3 (b).

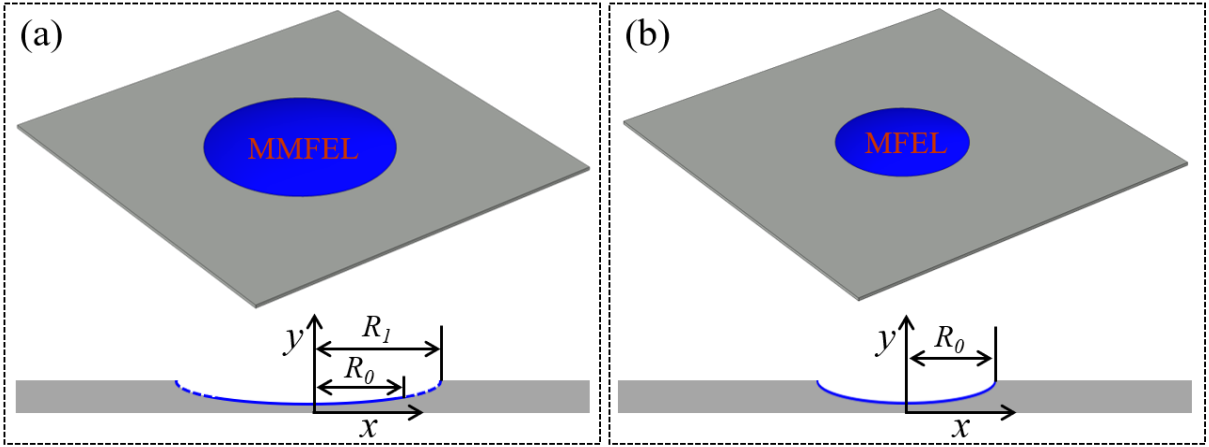

**Figure 3:** Structural lens. (a) MMFEL and (b) MFEL.

## 3. Numerical Studies

In this study, the proposed lens for super-resolution imaging on a thin aluminium plate is investigated by numerical simulations utilizing the finite element method. For the plate, the following material properties are used: Young's modulus $E = 70$ GPa, Poisson's ratio $\upsilon = 0.33$, and density $\rho = 2700$ kg/m$^3$. The plate dimension is 650 mm × 650 mm× 5 mm. Simulations in both time and frequency domain are carried out by the COMSOL software. To minimize reflections from edges of the plate, perfectly matched layers (PML) are used. No structural damping is included in the simulations in order to evaluate the performance of the lens for broadband super-resolution focusing without the influence of structural damping. A harmonic signal with a central frequency of 5 kHz is used as the excitation signal at a distance of 100 mm from the center of the lens.

Firstly, to show the super-resolution imaging performance of the structural MMFEL at each moment, transient analyses are performed numerically. A circular wave is generated by a point source located at $(x, y) = (-0.1$ m, $0$ m$)$. A 3-cycle tone burst signal with a central frequency of 5 kHz with a displacement amplitude of 1 mm is used as the excitation signal. Figure 4 depicts the obtained waveforms at time intervals $t = 0.2, 0.5, 0.7$, and $0.9$ millisecond for both MMFEL and MFEL. At 0.2 millisecond, circular waveforms are generated and propagate in a forward direction. At 0.5 millisecond, the waves propagate within the GRIN lens. This results in flattened wave fronts. At 0.7 millisecond, the plane waves become narrower and more focused for both of the lenses. The maximum displacement amplitude of MMFEL is 2 mm, while the maximum displacement amplitude of MFEL is 1.5 mm, which validates the energy density of MMFEL is higher compared with the one of MFEL. At 0.9 millisecond, the focused waves radiate outside in the propagation direction.

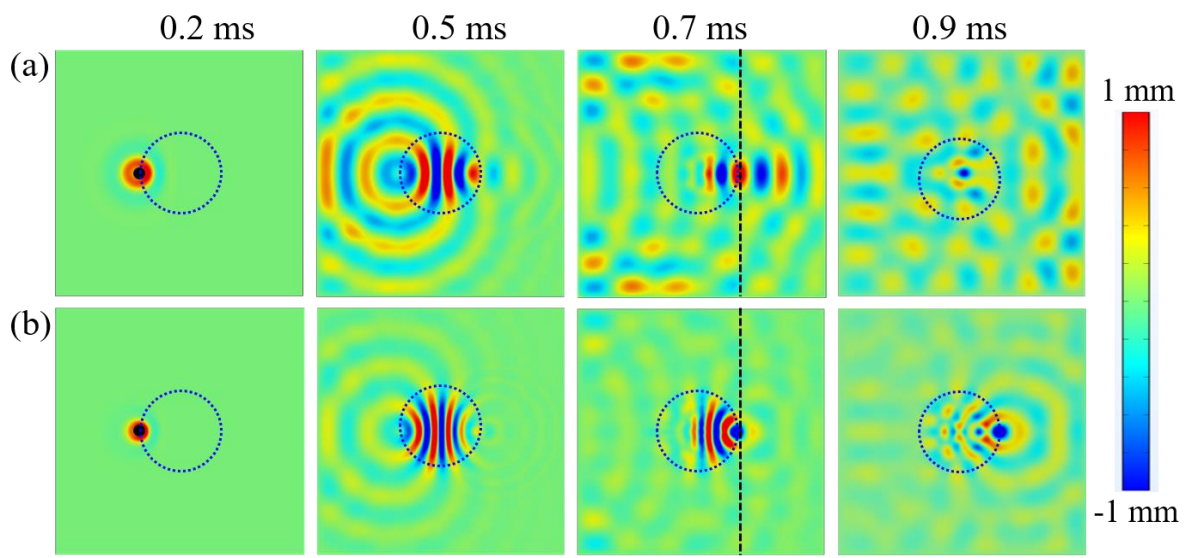

**Figure 4: Transient response of structural lens at time instants of $t$ = 0.2 ms, $t$ = 0.5 ms, $t$ = 0.7 ms, and $t$ = 0.9 ms. (a) and (b) are numerical simulations obtained for MFEL and MMFEL, respectively.**

To objectively evaluate the super-resolution imaging capabilities of the structural lens, we measured the peak value of the displacement amplitude field distributions of both lenses along the $y$-axis passing through the focal positions (vertical black dotted line in Figure 4), as shown in Figure 5(a). In addition, the full width at half maximum (FWHM) of the subwavelength imaging is obtained by measuring the amplitude along the $y$-direction. The full FWHM is the distance between two points in the $y$-direction where the shifts reach more than half of their peak value. The obtained results are shown in Figure 5(b) over a wide frequency range of 5-10 kHz. For all the simulated frequency range, the obtained FWHM are around $0.2\lambda$ for MMFEL, while the obtained FWHM are around $0.5\lambda$ for MFEL. These results prove that the MMFEL are capable of producing a super-resolution imaging compared with traditional MFEL.

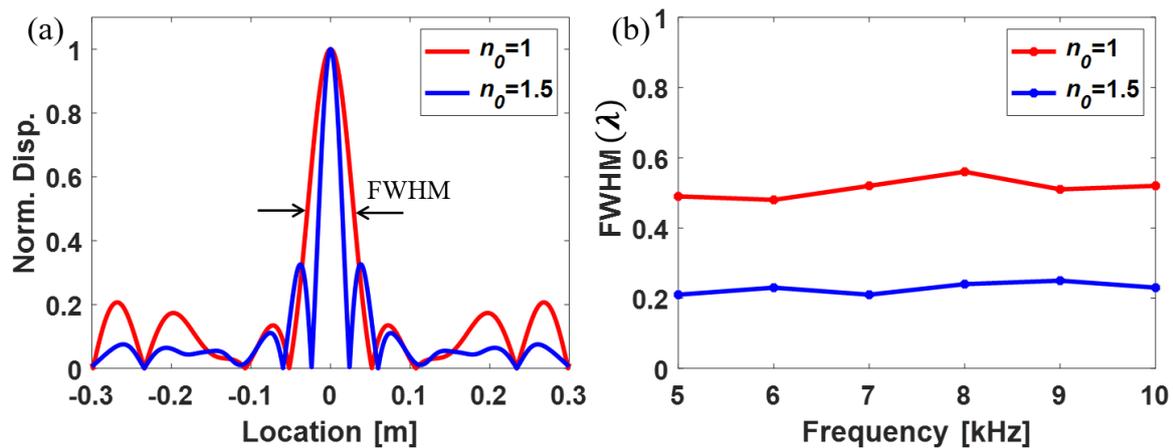

**Figure 5: Numerical simulations results. (a) The normalized displacement of the lens along the *y* axis at maximum amplitude spots, and (b) the obtained FWHM for different frequencies.**

In addition, to verify the effectiveness of super-resolution imaging for multiple excitation sources, finite element simulations are performed. Here, two point sources excited at $r = 0.1$ m from the center of lenses with angular distance $\alpha = 30°$ at the frequency of 5 kHz. The obtained waveforms are provided in Figure 6, which demonstrates that a super-resolution imaging can be achieved using MMFEL. Similarly, the maximum displacement amplitude field distributions of both lenses along the *y*-axis at the tangent of the lenses (represented by the vertical black dotted line in Figure 6 show that the two acoustic sources can be differentiated by MMFEL compared to MFEL (Figure 7).

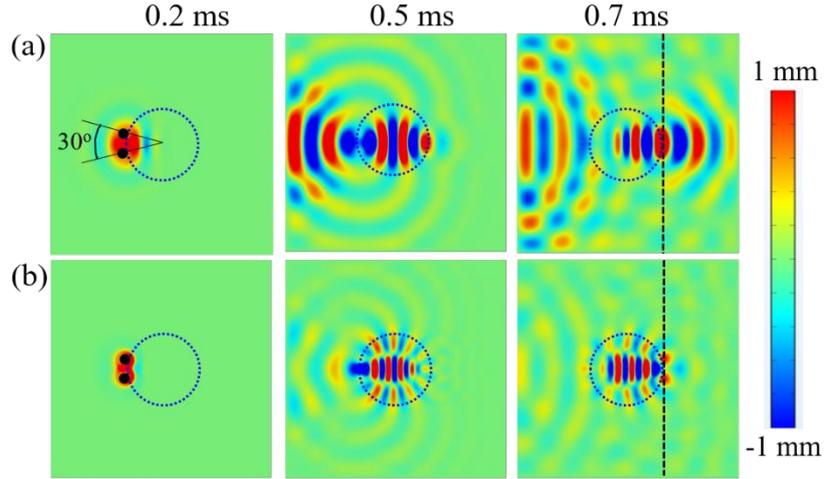

**Figure 6: Numerical simulations with two excitation sources for (a) MFEL and (b) MMFEL, respectively.**

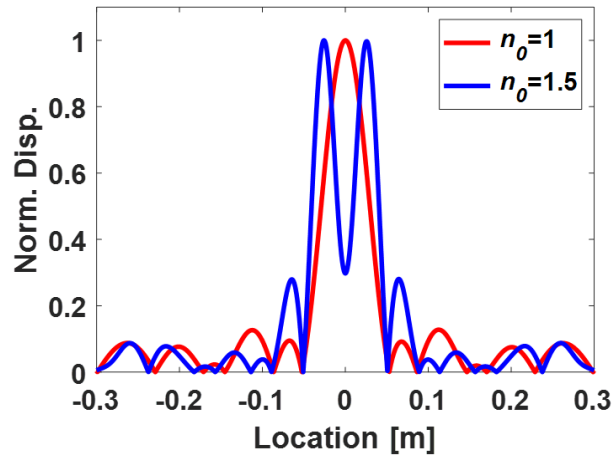

**Figure 7: Numerical simulations of the normalized displacement along the tangent of both MMFEL and MFEL.**

## 4. Experimental Studies

The lens's performance is also examined experimentally. It is manufactured through Computer Numerical Control machining on an aluminium plate with dimension of 650 mm × 650 mm × 4 mm, which matches that of the numerical simulations, depicted in Figure 8(a). Absorbing layers are utilized on the plate boundaries to diminish any reflections. The signal source for the three-cycle sound burst, with a central frequency of $f$ = 5 kHz, is generated using RIGOL DG4062 wave generator. To enhance the signal-to-noise ratio, amplification of the three-cycle tone burst is achieved using a power amplifier. Piezoelectric patches connected to the power amplifier are bonded to the plate for excitations. A Polytec NLV-2500 laser vibrometer mounted on the two-axis motorized stage is utilized as the receiving probe to measure the amplitude and phase of the out-of-plane displacement field. As the two-axis motorized stage moves incrementally, an oscilloscope (PicoScope 4000 series) captures and archives transmitted out-of-plane displacement fields data onto a computer. The corresponding scanning zone covers 100 × 100 mm². Furthermore, utilizing the input signal, wave fields of adjacent frequencies can be examined, rendering possible the acquisition of broadband frequency wave fields from a single measurement.

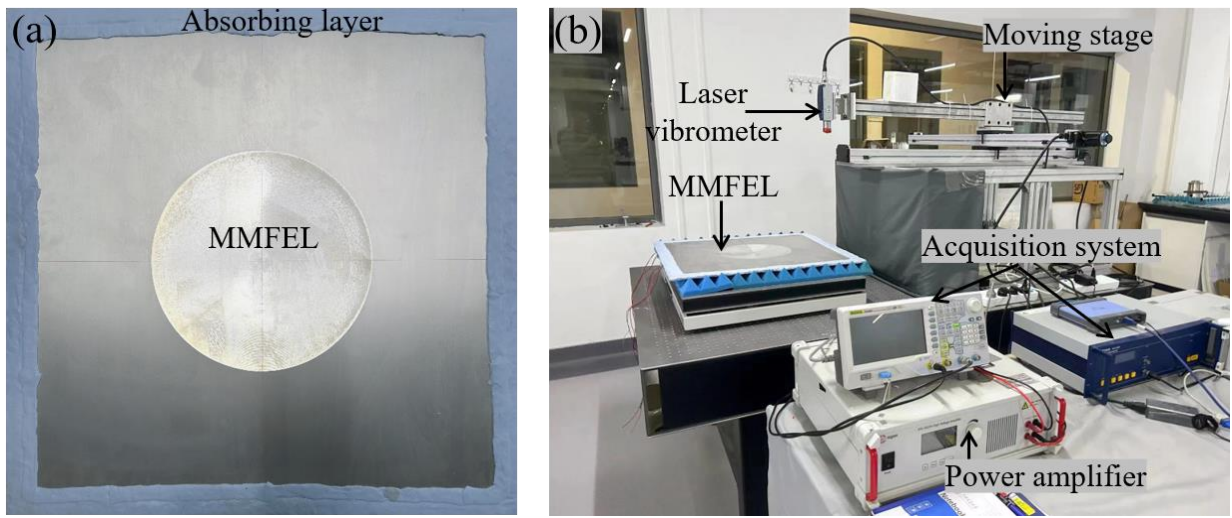

**Figure 8: Experimental setup. (a) Fabricated MMFEL lens. (b) Photo of the experimental setup. A laser vibrometer is utilized to measure the complete wave propagation within the scanned region.**

One circular patch of piezoelectric material with a radius of 10 mm is bonded to the manufactured plate at $r = 0.1$ m from the center of the lens, to serve as a point source. The coordinate of the piezoelectric patch agrees with the simulation in Figure 4(b), as can be observed from Figure 9(a). Additionally, two circular piezoelectric patches, each with a radius of 10 mm, are bonded to the manufactured plate at $r = 0.1$ m from the center of the lens. These patches are placed with an angular distance of $\alpha = 30^o$, as shown in Figure 9(d). Prior to this, we independently excited each piezoelectric patch to verify the deflection angles at different locations. Our measurements show the vertical displacement along the $y$-axis and field distribution, which are depicted in Figure 9(b)–(c) and Figure 9(e)–(f), respectively. The results clearly demonstrate the strong agreement between the captured super resolution images and their numerical counterparts. Consequently, the proposed lens for super-resolution imaging is undoubtedly validated for its effectiveness.

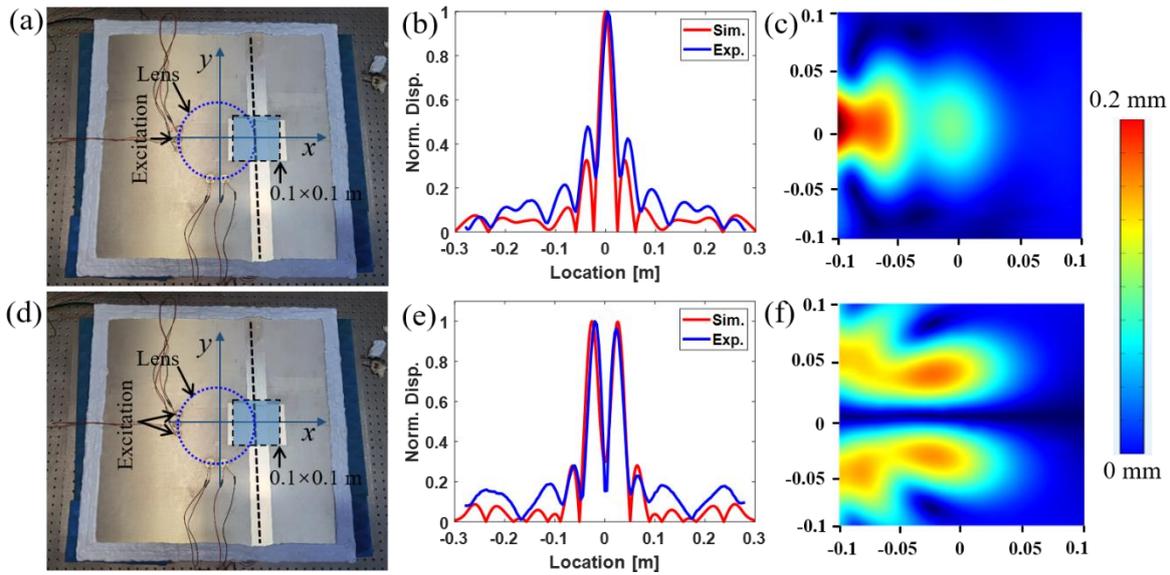

**Figure 9: (a) Experimental setup for one piezoelectric patch excitation. (b) Normalized displacement along $y$ axis. (c) Displacement amplitude field in the transparent blue area. (d) Experimental setup for two piezoelectric patches excitation. (e) Normalized displacement along $y$ axis. (f) Displacement amplitude field in the transparent blue area.**

In addition, the super-resolution imaging's FWHM is obtained from displacement amplitude along the $y$ direction. The outcomes are presented in Figure 10, covering a broad frequency range of 5-10 kHz. The experimental data show a consistent result with the numerical simulations, which demonstrate that the MMFEL can produce a super-resolution imaging in a broad frequency range.

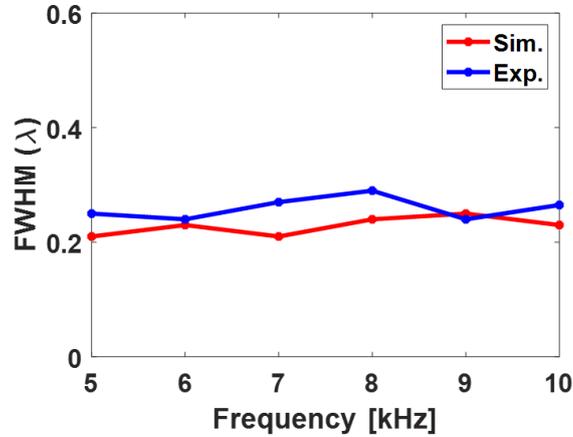

**Figure 10: Comparison of the FWHM between numerical and experimental results for different frequencies.**

## 5. Discussion

The characteristics of the proposed lens was analysed both numerically and experimentally in the previous sections. In this section, the performance of super-resolution imaging based on the MMFEL is discussed.

For one-point source excitation, the maximum displacement amplitude of MMFEL is 2 mm, while the maximum displacement amplitude of MFEL is 1.5 mm, which validates the energy density of the focal spot of MMFEL is higher compared with the one of MFEL. In addition, the obtained FWHM are around $0.2\lambda$ for MMFEL over a wide frequency range of 5-10 kHz, while the obtained FWHM are around $0.5\lambda$ for MFEL in the same frequency range. Similarly, for two-point source excitations, the maximum displacement amplitude field distribution of MMFEL has two peak values, while the maximum displacement amplitude field distribution of MFEL only has one peak value, which shows that the two acoustic sources can be differentiated by MMFEL compared to MFEL. These results prove that the MMFEL are capable of producing a super-resolution imaging compared with traditional MFEL.

Note that there are slight discrepancies between the simulation results and the experimental measurements. These can be attributed to the following reasons: i) The material properties of the lens used in experiments may not perfectly match with those used in numerical simulations. ii) The adhesive layers used to bond the piezo transducers with the structure in the experimental study were not considered in the numerical simulations, which can generate the inconsistencies between the numerical and experimental results. iii) The imperfection of the fabricated structural lens may affect the performance of the lens. iv) The excitation used in the experiments was not a perfect point source, while a perfect point source excitation was used in the simulations. However, the main characteristics of experimental measurements are consistent with the numerical results.

## 6. Conclusions

In conclusion, this study investigates a new design for a structural lens with a gradient refraction index, inspired by Maxwell's fish-eye lens. The lens is constructed using a circular structure with varying thickness, resulting in a continuous change of refractive index. This design enables the broadband, super-resolution imaging of flexural waves. Both numerical simulations and experimental tests are performed to explore the lens' broadband super-resolution imaging capabilities. The study indicates that the structural lens can attain a FWHM of approximately $0.2\lambda$ in the range of 5-10 kHz. This could potentially enhance vibration-based energy harvesting and non-destructive evaluations.

**Acknowledgments**

The authors sincerely acknowledge the financial support of Anhui Provincial Natural Science Foundation (Grant No. JZ2023AKZR0583), the Fundamental Research Funds for the Central Universities (Grant No. JZ2023HGTB0215), and the National Natural Science Foundation of China (Grant No. 12172271).

**Conflict of Interest**

The authors declare no conflict of interest.

**Data Access Statement**

The corresponding author can provide the data supporting the findings of this study upon reasonable request.